\documentclass{article}

\usepackage{arxiv}

\usepackage[utf8]{inputenc}
\usepackage[T1]{fontenc}
\usepackage{url}
\usepackage{booktabs}
\usepackage{nicefrac}
\usepackage{microtype}
\usepackage{graphicx}
\usepackage[numbers,sort&compress]{natbib}
\usepackage{doi}
\usepackage{tabularx}

\usepackage{amsmath}
\usepackage{amssymb}
\usepackage{amsfonts}
\usepackage{bm}
\usepackage{mathtools}
\usepackage{amsthm}
\usepackage{algorithm}
\usepackage{algpseudocode}
\usepackage{xcolor}
\usepackage{svg}
\usepackage{tikz}
\usepackage{wrapfig}
\usepackage{enumitem}
\usepackage{array}
\usepackage{threeparttable}
\usepackage{pgfplots}
\usepackage{pgfplotstable}
\usepackage{needspace}
\usepackage{multirow}
\usepackage{subcaption}

\usetikzlibrary{
    shapes.geometric,
    arrows.meta,
    positioning,
    fit,
    backgrounds,
    calc
}

\usepgfplotslibrary{groupplots}
\usepgfplotslibrary{statistics}
\pgfplotsset{compat=1.18}

\usepackage{hyperref}

\definecolor{darayuscolor}{HTML}{D62728}
\definecolor{ahmedcolor}{HTML}{1F77B4}
\definecolor{srijeetcolor}{HTML}{2CA02C}
\definecolor{subhomoycolor}{HTML}{9467BD}
\definecolor{abubakrcolor}{HTML}{FF7F0E}
\definecolor{arnocolor}{HTML}{17BECF}
\definecolor{vinaycolor}{HTML}{E377C2}
\definecolor{souravcolor}{HTML}{8C564B}

\definecolor{c4}{RGB}{0,114,178}
\definecolor{c8}{RGB}{230,159,0}
\definecolor{c16}{RGB}{0,158,115}
\definecolor{c32}{RGB}{204,121,167}
\definecolor{c64}{RGB}{213,94,0}

\colorlet{c1}{blue!70}
\colorlet{c2}{orange!90!black}
\colorlet{c3}{teal!75!black}

\definecolor{realColor}{HTML}{0F766E}
\definecolor{realSoft}{HTML}{E6F4F1}
\definecolor{simColor}{HTML}{1D4ED8}
\definecolor{simSoft}{HTML}{EEF2FF}
\definecolor{benchColor}{HTML}{475569}
\definecolor{calibFill}{HTML}{FBBF24}
\definecolor{calibLine}{HTML}{D97706}
\definecolor{fbColor}{HTML}{E11D48}

\newcommand{\system}{\textsc{CALIPER}}

\newcommand{\circletag}[1]{%
  \tikz[baseline=(n.base)]{
    \node[
      circle,
      fill=gray!85,
      text=white,
      font=\sffamily\bfseries\fontsize{8.5}{9}\selectfont,
      minimum size=1em,
      inner sep=0pt
    ] (n) {#1};
  }%
}

\newcounter{keytakeaway}

\newcommand{\keytakeaway}[1]{%
    \refstepcounter{keytakeaway}%
    \par\medskip
    \noindent
    \begingroup
    \setlength{\fboxsep}{6pt}%
    \fcolorbox{realColor}{realSoft}{%
        \parbox{\dimexpr\linewidth-2\fboxsep-2\fboxrule\relax}{%
            \textbf{\textcolor{realColor}{%
                Key takeaway \thekeytakeaway:}} #1%
        }%
    }%
    \endgroup
    \par\medskip
}

\title{\system: Calibrating HPC/AI Network Simulators}

\date{}

\author{
Seyyidahmed Lahmer \\
imec \\
Leuven, Belgium \\
\texttt{seyyidahmed.lahmer@imec.be}
\And
Daniel Seibel \\
Forschungszentrum J\"ulich (FZJ) \\
J\"ulich, Germany \\
\texttt{d.seibel@fz-juelich.de}
\And
Srijeet Basu \\
imec \\
Leuven, Belgium \\
\texttt{Srijeet.Basu@imec.be}
\And
Arno Troch \\
University of Antwerp -- imec \\
Antwerp, Belgium \\
\texttt{Arno.Troch@imec.be}
\And
Jayesh Badwaik \\
Forschungszentrum J\"ulich (FZJ) \\
J\"ulich, Germany \\
\texttt{j.badwaik@fz-juelich.de}
\And
Subhomoy Bhattacharyya \\
imec \\
Leuven, Belgium \\
\texttt{Subhomoy.Bhattacharyya@imec.be}
\And
Abubakr Nada \\
imec \\
Leuven, Belgium \\
\texttt{Abubakr.Nada@imec.be}
\And
Debjyoti Bhattacharjee \\
imec \\
Leuven, Belgium \\
\texttt{Debjyoti.Bhattacharjee@imec.be}
\And
Vinay Kumar Baapanapalli Yadaiah \\
imec \\
Leuven, Belgium \\
\texttt{Vinay.Kumar.BaapanapalliYadaiah@imec.be}
\And
Michael Peeters \\
imec \\
Leuven, Belgium \\
\texttt{michael.peeters@imec.be}
\And
Andreas Herten \\
Forschungszentrum J\"ulich (FZJ) \\
J\"ulich, Germany \\
\texttt{a.herten@fz-juelich.de}
\And
Darayus Adil Patel \\
imec \\
Leuven, Belgium \\
\texttt{DarayusAdil.Patel@imec.be}
}

\renewcommand{\shorttitle}{CALIPER: HPC/AI Network Simulator Calibration}

\hypersetup{
    pdftitle={CALIPER: Calibrating HPC/AI Network Simulators},
    pdfauthor={
        Seyyidahmed Lahmer,
        Daniel Seibel,
        Srijeet Basu,
        Arno Troch,
        Jayesh Badwaik,
        Subhomoy Bhattacharyya,
        Abubakr Nada,
        Vinay Kumar Baapanapalli Yadaiah,
        Michael Peeters,
        Andreas Herten,
        Darayus Adil Patel
    },
    pdfkeywords={
        High Performance Computing,
        Artificial Intelligence,
        Network Simulation,
        Simulator Calibration,
        Collective Communication,
        Network Validation
    }
}

\begin{document}

\maketitle

\begin{abstract}

Network simulation plays an important role in HPC/AI system research by enabling exploration of network architectures, communication strategies, and system configurations at scales that may be impractical to study experimentally. Existing approaches to network simulator calibration and validation lack a principled methodology that combines parameter alignment, constrained inference, and independent validation for collective-centric HPC/AI networks.
We present \system, a methodology for systematic calibration and validation of HPC/AI network simulators. \system{} distinguishes directly observable system parameters from auxiliary parameters which are configured via calibration and constrains the latter to feasible domains consistent with
the target system.
We evaluate \system's hierarchical, scale-aware calibration and its effects on parameter identifiability and generalization using collective communication benchmarks on a reference system with up to 64-nodes at the J\"ulich Supercomputing Centre. Validation results show that the calibrated parameter configurations identified by \system{} predict system performance with less than 5\% mean errors across scale-up and scale-out network domains. \system{} demonstrates that hierarchical, constrained calibration can produce simulator configurations whose predictive accuracy persists beyond the observations used for fitting.

\end{abstract}



\section{Introduction}

The continued scaling of artificial intelligence (AI) and high-performance computing (HPC) workloads is transforming network performance from a secondary consideration into a first-order system-design constraint~\cite{google-ici, nvidia-nvlink,ualink}. Contemporary systems distribute computation and model state across large numbers of compute nodes, making application performance dependent on the interplay between computation and communication. Collective communication primitives~\cite{Chan07CollectiveCommunication} are widely used to exchange data among compute nodes in workloads such as distributed training, model serving, and other communication-intensive applications. Consequently, network design is a key consideration in large-scale systems and must be evaluated under realistic workloads and configurations. However, direct hardware experimentation alone cannot support this evaluation. Representative systems are costly to construct and operate whereas available production installations may be unavailable for controlled experiments. Therefore, simulation provides a practical means to evaluate network architectures including implementation of collective algorithms, communication policies, etc.~\cite{rodrigues11sst,ns3,agarwal09garnet,jiang13noc}. It also provides repeatable and configurable experimental environments in which individual design variables can be isolated~\cite{casanova14simgrid, velho13tcp}. Because measurements on real HPC systems consume scarce compute and network resources, one of the requirements to have an effective methodology to configure and evaluate a network simulator is that it must be based on a deliberately selected, compact benchmark set.

Modern HPC/AI simulation studies increasingly report close correspondence with real systems, which establishes that accurate and scalable prediction is achievable~\cite{moolchandani23amped,kundu24distributedllm,won23astrasim,wang25simai,gui25parallel,shen25atlahs}. 
 However, the methodological path from system measurement to calibrated configuration is not described consistently across the literature~\cite{mcdonald24calibration,davoudzade24EnhancingNS}. {\em Calibration} determines parameter values that reduce misalignment between simulation and selected reference measurements, whereas {\em validation} evaluates whether the resulting model provides an acceptable level of accuracy for its intended use and domain of applicability.  This distinction is consequential: alignment with the measurements used for calibration demonstrates fit to the calibration data, but does not establish predictive validity. A specific instance of parameters of the simulator may achieve respectable accuracy for one benchmark while failing for a different workload or system scale. Calibration accuracy must consequently be evaluated along several generalization dimensions: unseen message sizes, held-out collectives, alternative workload configurations, and larger system scales~\cite{sargent20VnV,mehta16SimulationCredibility}.

Simulator parameters can be broadly classified into two categories namely, {\em direct parameters} that are set based on the target hardware information (e.g. the number of compute nodes) and {\em auxiliary parameters} cannot be directly associated with target hardware or are specific to the simulator abstraction (e.g. link latency). During calibration, {\em direct parameters} must not be tuned to reduce simulation error, as doing so would effectively model a non-existent system. {\em Auxiliary parameters} may be tuned during the calibration process and therefore define the calibration search space. Furthermore, the calibrated simulator must be explicitly validated under conditions systematically excluded from calibration, such as different collectives, message sizes, or system scales to assess its predictive capability on unseen workload scenarios.  
Existing approaches (Refer Section~\ref{sec:related_work}) to tune and evaluate network simulators typically do not jointly address simulator independence, generalization, constrained parameter inference, and system validation as shown in Table~\ref{tab:related-work-design-goals}.

In this paper, we introduce \system{} methodology for calibrating  network simulators for large scale HPC/AI systems. Specifically, this paper makes the following contributions.
\begin{itemize}
    \item We propose \system{}, a methodology that combines explicit parameter classification, bounded optimization, and independent validation to calibrate and evaluate network simulators for large scale AI/HPC systems.

    \item We apply \system{} for calibrating the SST simulator~\cite{rodrigues11sst} to simulate state-of-the-art network configurations~(with up-to 64 GPUs) from Jülich Supercomputing Centre.

    \item We show that \system’s hierarchical, scale-aware calibration produces simulator configurations that achieve less than 5\% mean completion-time error validated across the evaluated scale-up and scale-out network domains.


\end{itemize}
The reported results demonstrate that employing the proposed \system{} methodology enables a simulator configuration to retain its accuracy beyond the calibration conditions across collective measurements spanning message sizes, communicator scales, and physical placements.\\
The rest of this paper is organized as follows. Section~\ref{sec:background} provides the necessary background. Section~\ref{sec:sys_model} presents the \system{} model and its design. Section~\ref{sec:evaluation} presents the experimental setup and evaluation results. Section~\ref{sec:discussion} discusses the potential extensions of \system{} that can be undertaken in future work. Section~\ref{sec:related_work} reviews related work, while Section~\ref{sec:conclusion} concludes the paper.

\begin{table}[t]
\centering
\caption{Comparison of state-of-the-art calibration works with \system.}
\label{tab:related-work-design-goals}
\small
\begin{tabular}{l*{4}{>{\centering\arraybackslash}m{6.0em}}}
\toprule
& \textbf{\shortstack{Simulator\\agnostic}}
& \textbf{\shortstack{Generalization}}
& \textbf{\shortstack{Constrained\\calibration}}
& \textbf{\shortstack{System\\validation}} \\
\midrule
Mehta et al. (2016)~\cite{mehta16SimulationCredibility}     & \checkmark & \checkmark & -- & \checkmark \\
Sargent (2020)~\cite{sargent20VnV}         & \checkmark & \checkmark & -- &  \checkmark \\
McDonald et al. (2024)~\cite{mcdonald24calibration}  & \checkmark & -- & \checkmark &  -- \\
Davoudzade et al. (2024)~\cite{davoudzade24EnhancingNS} & -- & \checkmark & -- & \checkmark \\
\midrule
\textbf{\system}                 & $\boldsymbol{\checkmark}$ & $\boldsymbol{\checkmark}$ & $\boldsymbol{\checkmark}$ & $\boldsymbol{\checkmark}$ \\
\bottomrule
\end{tabular}
\end{table}

\section{Background}
\label{sec:background}

Contemporary standalone network simulators as well as distributed AI system simulators compare predicted communication or application performance with measurements from accelerator clusters and supercomputing systems, in some cases at large system scales~\cite{hoefler10loggopsim,casanova01simgrid,casanova14simgrid,casanova25simgrid,handleyndhoma,degomme17smpi,won23astrasim,isaev23calculon,moolchandani23amped,kundu24distributedllm,wang25simai,gui25parallel,shen25atlahs}. These studies show that useful predictive accuracy can be achieved using a variety of modeling strategies at different abstraction levels, from analytical models to cycle-accurate simulation, each offering a distinct balance between execution speed, scalability, and fidelity of network representation.

Table~\ref{tab:network-simulator-taxonomy} summarizes the taxonomy of simulator types and highlights their respective trade-offs. Analytical models~\cite{hoefler10loggopsim,lu17flexflow,won23astrasim,isaev23calculon,moolchandani23amped,ardalani24deepflow,kundu24distributedllm,shen25atlahs} prioritize speed and scalability at the expense of modelling granularity. Flow/event-based simulators~\cite{ns3,casanova01simgrid,casanova14simgrid,degomme17smpi,handleyndhoma,li19hpcc,cho24llmservingsim,wang25simai,gui25parallel,casanova25simgrid} explicitly model communication operations and resource sharing while abstracting some packet- or cycle-level behaviour. Cycle-accurate network simulators~\cite{rodrigues11sst,agarwal09garnet,jiang13noc} model router pipelines, buffers, arbitration, links, and flit movement at or near individual clock cycles. However, abstraction level alone does not determine predictive accuracy. A more detailed simulator can represent mechanisms that a coarser model omits, but its predictions remain meaningful only if the simulated workload and model parameters correspond to the target system. Conversely, a compact model can provide accurate predictions within an operating regime in which its assumptions and parameters are appropriate. Predictive accuracy is therefore not an intrinsic property of a simulator category but depends on the calibration and validation of a specific model for its target system and workload.

\begin{table}[t]
\centering
\caption{Taxonomy of network simulator types.}
\label{tab:network-simulator-taxonomy}
\small

\resizebox{\linewidth}{!}{%
\begin{tabular}{lccc}
\toprule
\textbf{} & \textbf{Analytical} & \textbf{Flow/Event-based} & \textbf{Cycle-accurate} \\
\midrule

\textbf{Approach}
& Mathematical models
& Discrete-event, packet-based
& Cycle/flit-level \\

\textbf{Speed}
& High
& Medium
& Low \\

\textbf{Scalability}
& High
& Medium
& Low \\

\textbf{Strengths}
& Fast, interpretable
& Speed/accuracy balance
& Accuracy, realism \\

\textbf{Weaknesses}
& High abstraction level
& Needs calibration
& Costly at scale \\

\textbf{Simulators}
& {\small \cite{hoefler10loggopsim,lu17flexflow,won23astrasim,isaev23calculon,moolchandani23amped,ardalani24deepflow,kundu24distributedllm,shen25atlahs}}
& {\small \cite{ns3,casanova01simgrid,casanova14simgrid,degomme17smpi,handleyndhoma,li19hpcc,cho24llmservingsim,wang25simai,gui25parallel,casanova25simgrid}}
& {\small \cite{rodrigues11sst,agarwal09garnet,jiang13noc}} \\

\bottomrule
\end{tabular}%
}

\end{table}

Validation is increasingly reported across both distributed AI simulators~\cite{lu17flexflow,won23astrasim,isaev23calculon,moolchandani23amped,ardalani24deepflow,kundu24distributedllm,cho24llmservingsim,lee25forecasting,wang25simai,gui25parallel,shen25atlahs,svedas2026evaluating} and standalone network simulators~\cite{casanova01simgrid,casanova14simgrid,casanova25simgrid,rodrigues11sst,ns3,agarwal09garnet,hoefler10loggopsim,jiang13noc,degomme17smpi,handleyndhoma,li19hpcc}. However, calibration procedures are often only partially described or omitted altogether. This lack of information limits the ability to assess, reproduce, and transfer simulation models across system configurations, motivating the need for more systematic calibration methodologies. There also exists considerable diversity in how calibration is performed~\cite{handleyndhoma,won23astrasim,isaev23calculon,lee25forecasting,wang25simai,gui25parallel,shen25atlahs}. Consequently, reported prediction accuracy on a selected validation set does not necessarily imply that the calibrated model will retain its predictive fidelity when applied to different calibration parameters, workloads or system configurations. The resulting gap is therefore not an absence of calibration or validation, but rather the fragmented methodological treatment of how calibration should be designed, executed, and evaluated so that the resulting model is both accurate and demonstrably transferable beyond the observations used to fit it.

A fundamental limitation of calibration based solely on quality of fit is that different parameter configurations can explain the same finite set of measurements, and low fitting error does not imply that the inferred parameters capture reusable system behavior. Automated search can make this process faster and reduce the effort of manual tuning~\cite{mcdonald24calibration}, but optimization alone cannot determine whether a low-error configuration is physically meaningful or whether its apparent accuracy results from compensating errors. This makes simulator configuration consequential. Some simulator inputs correspond directly to properties of the target system, such as topology, nominal link rate, job, or routing constraints. These quantities should, whenever possible, be obtained from the hardware platform and held fixed. Other parameters represent effective quantities that cannot be measured directly at the simulator interface and must instead be inferred from observations. Finally, some real-system behavior may not be modeled by the simulator at all. These cases require different treatment, although a sufficiently flexible optimizer can make them appear indistinguishable if every performance-sensitive parameter is allowed to vary. The methodological issue is therefore not whether simulation parameters can be adjusted to improve error;\footnote{ error, loss, and agreement are used interchangeably throughout the manuscript} with measurements. The harder question is how to determine \emph{which} quantities should be adjusted, \emph{under what constraints}, and \emph{what evidence} is sufficient to show that the resulting configuration remains predictive beyond the measurements used to obtain it. 

\system{} addresses this gap by separating three sources of simulator discrepancy that tuning would otherwise conflate. First, directly observable properties and execution choices are aligned between the target system and the simulation. Second, only explicitly uncertain auxiliary parameters are exposed to calibration. Third, the resulting configuration is frozen and evaluated on structured, held-out workloads~\cite{sargent20VnV} and system scales. Residual discrepancies that persist after these stages are not automatically absorbed through further tuning; instead, they provide evidence of workload-representation, implementation, or model fidelity limitations. This separation makes calibration interpretable while providing stronger confidence that the calibrated simulator can generalize beyond calibrated dataset.

\section{\system}
\label{sec:sys_model}

\begin{figure*}[t]
    \centering
    \includegraphics[width=0.95\textwidth]{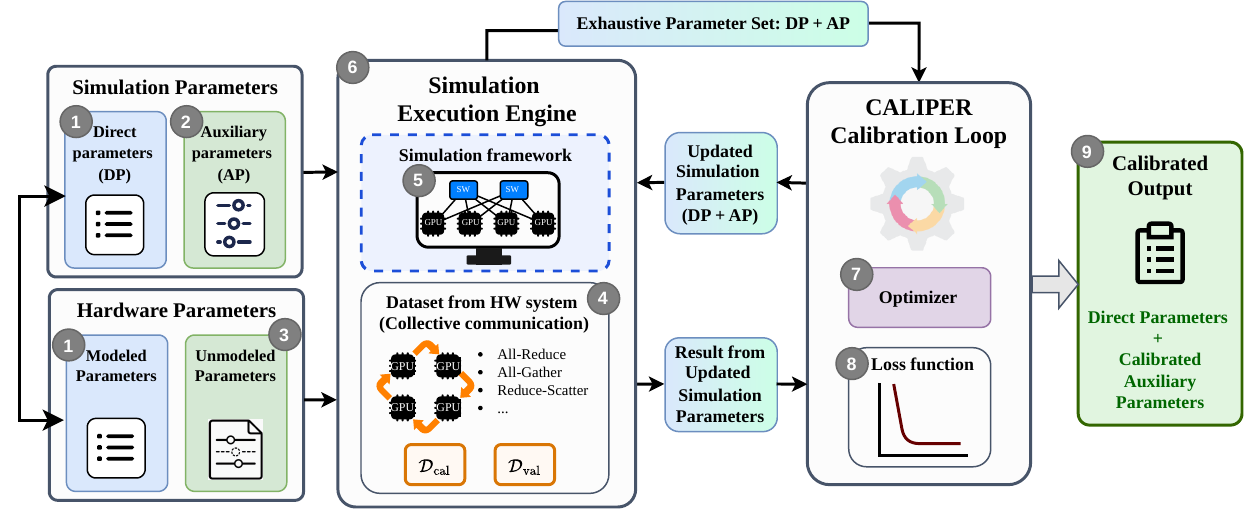}
    \caption{\system{} overview. Direct parameters (DP) remain aligned with the target system, while auxiliary parameters (AP) are iteratively optimized against the calibration dataset $\mathcal{D}_{\mathrm{cal}}$; the resulting configuration is fixed before evaluation on the validation dataset $\mathcal{D}_{\mathrm{val}}$}
    \label{fig:model}
\end{figure*}

In this section, we introduce the \system{} methodology, shown in Figure~\ref{fig:model}. \system{} partitions the simulator configuration into three parameter classes. \emph{Direct parameters}~(DP) ~\circletag{1}, correspond to observable or specified properties of the target system~\circletag{1}, such as GPU count, link bandwidth, topology, etc. and are aligned with the hardware configuration rather than optimized. \emph{Auxiliary parameters}~(AP)~\circletag{2}, are simulator-specific, non-directly measurable parameters whose values are inferred through calibration to reproduce target-system behavior. Auxiliary parameters lack one-to-one correspondence with the target system or exist primarily within the simulator abstraction; these constitute the calibration search space. System characteristics that are not represented in the simulator, i.e., parameters unmodeled in simulation~\circletag{3}, are classified as model gaps and require refinement of the underlying model or implementation rather than parameter adjustment.

\system{} executes equivalent collective communication benchmarks~\circletag{4} on the target system and simulation framework~\circletag{5}. The Simulation Execution Engine~\circletag{6} provides a simulator-agnostic interface for configuring and executing the target simulation framework and comparing its outputs against hardware measurements partitioned into calibration $\mathcal{D}_{\mathrm{cal}}$ and validation $\mathcal{D}_{\mathrm{val}}$ datasets. During calibration, direct parameters remain fixed at their hardware-aligned values, while the \system{} optimizer~\circletag{7} iteratively updates the auxiliary parameters based on the loss function~\circletag{8} between simulated results and $\mathcal{D}_{\mathrm{cal}}$. The resulting calibrated output~\circletag{9} combines the fixed direct parameters with the optimized auxiliary parameters.

Although we use collective communication~\cite{Chan07CollectiveCommunication} throughout the paper as a concrete example, the formulation is independent of a specific simulator, workload, or performance metric. Table~\ref{tab:system-model-notation} summarizes the notation used in the \system{} methodology elaborated in sub-sections ~\ref{subsec:experiments}--\ref{subsec:optimizer}.

\begin{table*}[t]
    \centering
    \caption{Notation used in the \system{} methodology.}
    \addtolength{\tabcolsep}{-0.4em}
    \footnotesize
    \renewcommand{\arraystretch}{1.10}
    \begin{tabular}{
        >{\raggedright\arraybackslash}p{0.15\textwidth}
        >{\centering\arraybackslash}p{0.17\textwidth}
        >{\raggedright\arraybackslash}p{0.60\textwidth}
    }
        \toprule
        \textbf{Category} & \textbf{Symbol} & \textbf{Meaning} \\
        \midrule

        System
        & $\mathcal{H}$
        & Target real system from which performance measurements are collected. \\

        System
        & $\mathcal{S}$
        & Simulator used to predict the behavior of the target system. \\

        Experiment
        & $\mathcal{X}$
        & Space of benchmarks and system configurations considered for
          calibration. \\

        Experiment
        & $x \in \mathcal{X}$
        & One experiment, including its workload and system configuration. \\

        Experiment
        & $(c,n,m,p,\tau)$
        & Collective operation, number of participating accelerators, message
          size, accelerator placement, and topology or system configuration. \\

        Observation
        & $y(x)$
        & Performance measured on the target system for experiment $x$.
          In this paper, it is collective completion time. \\

        Prediction
        & $\hat{y}(x;\phi,\theta)$
        & Simulator prediction for experiment $x$ under direct parameters
          $\phi$ and auxiliary parameters $\theta$. \\

        Parameter
        & $\phi$
        & Directly observable or confidently known simulator parameters.
          These parameters remain fixed during calibration. \\

        Parameter
        & $\theta=(\theta_1,\ldots,\theta_d)$
        & Vector of $d$ auxiliary parameters that the calibration procedure
          is allowed to modify. \\

        Parameter domain
        & $\Theta_i$
        & Feasible domain of auxiliary parameter $\theta_i$. \\

        Parameter domain
        & $\Theta$
        & Joint feasible parameter space,
          $\Theta=\Theta_1\times\cdots\times\Theta_d$. \\

        Dataset
        & $\mathcal{D}_{\mathrm{cal}}$
        & Measurements used to infer the auxiliary parameters. \\

        Dataset
        & $\mathcal{D}_{\mathrm{val}}$
        & Independent measurements used only to evaluate the calibrated
          configuration. \\

        Experiment group
        & $\mathcal{G}_{\mathrm{cal}}$
        & Set of experiment groups included in calibration. Within one group,
          the collective, communicator size, placement, and execution
          configuration remain fixed while message size varies. \\

        Experiment group
        & $\mathcal{D}_k$
        & Calibration observations belonging to experiment group $k$. \\

        Loss
        & $L_k(\theta)$
        & Root-mean-square completion-time error over the observations in
          experiment group $k$. \\

        Loss
        & $\mathcal{L}(\theta)$
        & Overall calibration objective. For multiple experiment groups, it is
          the maximum group-level loss. \\

        Result
        & $\theta^\star$
        & Auxiliary parameter vector that minimizes the calibration objective. \\

        Validation
        & $E_i(\theta^\star)$
        & Absolute relative prediction error for validation observation $i$. \\

        \bottomrule
    \end{tabular}
    \label{tab:system-model-notation}
\end{table*}

\subsection{Experiments and Metric Definition}
\label{subsec:experiments}

Let $\mathcal{H}$ denote the target system and $\mathcal{S}$ denote a simulator intended to model $\mathcal{H}$.
We define an experiment by a configuration
\begin{equation}
    x \in \mathcal{X},
\end{equation}
where $\mathcal{X}$ is the space of workloads and system configurations of interest. For collective communication, $x$ includes
\begin{equation}
    x =
    (c, n, m, p, \tau, \ldots),
\end{equation}
where $c$ is the collective operation, $n$ is the number of participating accelerators, $m$ is the message size, $p$ describes process or accelerator placement, and $\tau$ identifies the relevant topology or configuration. The formulation naturally extends to additional dimensions.

Executing $x$ on the target system produces a performance observation

\begin{equation}
\label{eq:hardware-observation}
    y(x) = \mathcal{H}(x).
\end{equation}

In this paper, $y(x)$ is the \emph{collective completion time}, defined as the elapsed time from the invocation of the collective operation until the last participating rank completes it.

We also report \emph{algorithmic bandwidth}, defined as

\begin{equation}
\label{eq:algorithmic-bandwidth}
    B_{\mathrm{alg}}(x)
    =
    \frac{m}{y(x)}.
\end{equation}

Following the definition used by \texttt{nccl-tests}~\cite{nccl_tests}, algorithmic bandwidth does not include a correction for the number of participating ranks. Factors that depend on $n$ are used only to derive NCCL bus bandwidth. For AllReduce, bus bandwidth is obtained by multiplying algorithmic bandwidth by $2(n-1)/n$. For AllGather and ReduceScatter, it is obtained by multiplying
algorithmic bandwidth by $(n-1)/n$. We report algorithmic bandwidth and do not use bus bandwidth as a separate evaluation metric.

\subsection{Simulator Parameters}
\label{subsec:simulator_parameters}
The simulator parameters are partitioned into two sets. The vector $\phi$ contains \emph{direct parameters}: quantities that can be mapped from the target system or experiment with sufficiently high confidence. They are fixed during calibration. The vector
\begin{equation}
    \theta = (\theta_1,\ldots,\theta_d)
\end{equation}
contains the \emph{auxiliary parameters} that \system{} is allowed to tune. Each auxiliary parameter belongs to a feasible domain $\Theta_i$, yielding
\begin{equation}
    \theta \in
    \Theta =
    \Theta_1 \times \cdots \times \Theta_d.
\end{equation}
The domains may contain continuous, integer, or categorical values. Bounds encode simulator constraints and prior engineering knowledge and prevent the optimizer from producing meaningless simulator configurations.

For a given simulator $\mathcal{S}$ parameters and input workload, we represent the corresponding performance predicted by the simulator as
\begin{equation}
    \hat{y}(x; \phi, \theta)
        = \mathcal{S}(x; \phi, \theta).
\end{equation}

\subsection{Calibration Dataset}
\label{subsec:calibration_dataset}

Let
\begin{equation}
    \mathcal{D}_{\mathrm{cal}}
      = \{(x_i,y_i)\}_{i=1}^{N}
\end{equation}
denote the calibration dataset, where $i$ indexes one measured experiment, $x_i$ is its experiment configuration, and 

\begin{equation}
    y_i = \mathcal{H}(x_i).
\end{equation}

is its measured collective completion time.

The calibration problem is

\begin{equation}
\label{eq:calibration}
    \min_{\theta\in\Theta}
    \mathcal{L}
    \left(
        \mathcal{D}_{\mathrm{cal}},
        \theta
    \right).
\end{equation}

Let $\mathcal{G}_{\mathrm{cal}}$ denote the set of experiment groups included in calibration. An experiment group fixes the collective operation, communicator size, placement, and execution configuration while varying message size.

For each group $g\in\mathcal{G}_{\mathrm{cal}}$, let $\mathcal{D}_g\subseteq\mathcal{D}_{\mathrm{cal}}$ denote its calibration
observations. \system{} calculates the root-mean-square completion-time error~(RMSE) 

\begin{equation}
\label{eq:curve-rmse}
    L_g(\theta)
    =
    \sqrt{
        \frac{1}{|\mathcal{D}_g|}
        \sum_{(x_i,y_i)\in\mathcal{D}_g}
        \left[
            \mathcal{S}(x_i;\phi,\theta)-y_i
        \right]^2
    }.
\end{equation}

When several groups (i.e., different collectives, placements, etc..) are calibrated jointly, \system{} uses the minimax objective

\begin{equation}
\label{eq:objective}
    \mathcal{L}(\theta)
    =
    \max_{g\in\mathcal{G}_{\mathrm{cal}}}
    L_g(\theta).
\end{equation}

The minimax objective, therefore, minimizes the error of the worst-fitting experiment group. It prevents a large improvement for one group from hiding poor accuracy for another group. When calibration contains one group, Equation~\eqref{eq:objective} reduces to that group's RMSE.

\subsection{Calibration Vs Validation}
\label{subsec:calibration_validation}

A low error on $\mathcal{D}_{\mathrm{cal}}$ only shows that a parameter configuration explains the observations used during fitting. It does not show that the configuration remains accurate for other observations.

The final validation dataset is denoted by $\mathcal{D}_{\mathrm{val}}$. It is not used for parameter fitting, wheres

\begin{equation}
\label{eq:validation-independence}
    \mathcal{D}_{\mathrm{val}} \cap \mathcal{D}_{\mathrm{cal}} = \emptyset.
\end{equation}

For each validation observation $(x_i,y_i)\in\mathcal{D}_{\mathrm{val}}$, we calculate the absolute relative completion-time error

\begin{equation}
\label{eq:validation}
    E_i
    =
    \left|
        \frac{
            \mathcal{S}(x_i;\phi,\theta^\star)-y_i
        }{
            y_i
        }
    \right|.
\end{equation}

We report the distribution, mean, and selected percentiles of $E_i$ over $\mathcal{D}_{\mathrm{val}}$. Algorithmic bandwidth Eq.~\ref{eq:algorithmic-bandwidth} is derived from the same completion-time observations and is used to interpret curve shape rather than as an independent validation metric.

\subsection{Experiment Representation}

Caliper represents each measured experiment as
\begin{equation}
    r_i =
    (x_i,q_i,y_i),
\end{equation}
where $x_i$ describes the logical experiment, $y_i$ is the measured
performance, and $q_i$ describes how the experiment was realized on the
target system.

The distinction between $x_i$ and $q_i$ is important. A logical workload does not always determine a unique execution. Software layers can select different algorithms, decompose work differently,
or expose different levels of parallelism. Two executions can therefore have the same high-level workload $x_i$ while producing different low-level behavior.

\system{} records the execution information required to reproduce the relevant behavior in simulation. We denote this execution information by $q_i$.

The exact contents of $q_i$ depend on the simulator and workload. \system{}  does not impose a specific representation.

\subsection{Experiment Mapping}
The experiment mapper converts a measured experiment into a simulator workload:
\begin{equation}
    z_i =
    \mathcal{M}(x_i,q_i).
\end{equation}

The goal of $\mathcal{M}$ is to preserve the properties of the real execution that affect the simulated metric. \system{}  therefore requires the simulator workload to be equivalent to the measured execution with respect to the modeled behavior:
\begin{equation}
    q_i^{\mathrm{real}}
    \equiv
    q_i^{\mathrm{sim}}.
\end{equation}

We refer to this property as \emph{execution equivalence}. Execution equivalence prevents the optimizer from correcting workload mismatch through unrelated simulator parameters. For example, if the simulated workload exposes less parallelism than the real execution, an optimizer could compensate by increasing the simulated bandwidth. Such a configuration may reduce calibration error while producing poor predictions for other workloads. \system{}  therefore constructs $z_i$ before optimization and keeps it fixed for all candidate parameter vectors.




\subsection{Calibration Loop}

Given a calibration set
\begin{equation}
    \mathcal{D}_{\mathrm{cal}}
    =
    \{r_i\}_{i=1}^{N},
\end{equation}
Caliper first constructs the corresponding simulator workloads
\begin{equation}
    z_i = \mathcal{M}(x_i,q_i).
\end{equation}

The optimizer then proposes a candidate auxiliary parameter vector
$\theta_b$.
Caliper evaluates the simulator for every calibration experiment:
\begin{equation}
    \hat{y}_{i,b}
    =
    \mathcal{S}
    \left(
        z_i;
        \phi_i,
        \theta_b
    \right).
\end{equation}

The predictions are compared with the measured targets using the calibration
loss from Section~\ref{sec:sys_model}:
\begin{equation}
    L_b =
    \mathcal{L}
    \left(
        \mathcal{D}_{\mathrm{cal}},
        \theta_b
    \right).
\end{equation}

The optimizer uses $L_b$ to select the next candidate.
After the search budget is exhausted, \system{}  returns
\begin{equation}
    \theta^\star
    =
    \arg\min_{\theta_b}
    L_b.
\end{equation}

Algorithm~\ref{alg:caliper} summarizes the process.

\begin{algorithm}[t]
\caption{Caliper calibration}
\label{alg:caliper}
\begin{algorithmic}[1]
\Require Calibration set $\mathcal{D}_{\mathrm{cal}}$
\Require Direct parameters $\phi$
\Require Auxiliary parameter space $\Theta$
\Require Search budget $B$
\ForAll{$r_i=(x_i,q_i,y_i)\in\mathcal{D}_{\mathrm{cal}}$}
    \State $z_i \gets \mathcal{M}(x_i,q_i)$
\EndFor

\State Initialize optimizer over $\Theta$

\For{$b=1,\ldots,B$}
    \State $\theta_b \gets \textsc{Propose}()$

    \ForAll{$r_i\in\mathcal{D}_{\mathrm{cal}}$}
        \State
        $\hat{y}_{i,b}
        \gets
        \mathcal{S}(z_i;\phi_i,\theta_b)$
    \EndFor

    \State
    $L_b
    \gets
    \mathcal{L}
    (\mathcal{D}_{\mathrm{cal}},\theta_b)$

    \State
    $\textsc{Observe}(\theta_b,L_b)$
\EndFor

\State \Return
$\theta^\star=\arg\min_{\theta_b}L_b$
\end{algorithmic}
\end{algorithm}

\subsection{Optimizer}
\label{subsec:optimizer}
\system{} treats the optimizer as a replaceable component. It exposes two operations:
\begin{itemize}
\item \textsc{Propose} returns a candidate parameter vector
\item \textsc{Observe} reports the corresponding objective value for the parameter vector.
\end{itemize} This separation is useful because simulator calibration is naturally a black-box optimization problem. Simulator evaluations may contain discrete events, integer parameters, categorical choices, and discontinuities. They also do not generally expose useful gradients. \system{} therefore does not depend on a specific optimization algorithm. The simulator and optimizer configuration used in the evaluation is described in Section~\ref{sec:simulator-configuration}. Evaluating one candidate parameter vector requires executing the simulator for all experiments in $\mathcal{D}_{\mathrm{cal}}$. These executions are independent and can be evaluated in parallel. system also caches simulator results using
\begin{equation}
    (z_i,\phi_i,\theta)
    \rightarrow
    \hat{y}_i.
\end{equation}

If the same configuration is requested again, \system{} reuses the cached result rather than invoking the simulator.

\section{Experimental Evaluation}
\label{sec:evaluation}
In this section, we evaluate the \system{} methodology to calibrate and validate a network simulator to faithfully represent target hardware platforms. 

\subsection{Target Hardware Platforms}
\begin{figure}[t] \centering 
\begin{subfigure}[t]{0.47\columnwidth} 
\centering 
\caption{Platform used for quantitative evaluation.} 
\label{tab:experimental-platform} 
\scriptsize \setlength{\tabcolsep}{3pt} 
\renewcommand{\arraystretch}{1.1} \resizebox{\linewidth}{!}{%
\begin{tabular}{ll} \toprule \textbf{Parameter} & \textbf{Value} \\ \midrule Platform & JUWELS Booster \\ GPUs & 4--64 \\ Accelerators/node & 4$\times$ A100-SXM4 \\ Scale-up topology & NVLink3 full mesh \\ Scale-out topology & Dragonfly+ \\ Interfaces/node & 4$\times$ ConnectX-6 HDR200 \\ 
\bottomrule 
\end{tabular}%
} 
\end{subfigure} \hfill 
\begin{subfigure}[t]{0.50\columnwidth} \centering 
\caption{Different placements on the JUWELS Booster Dragonfly+ topology.} 
\label{fig:placements} 
\includegraphics[width=\linewidth]{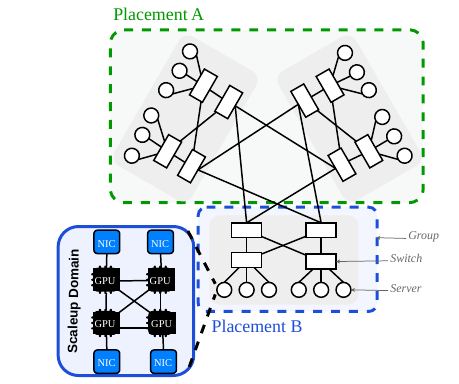} 
\end{subfigure} 
\caption{Experimental platform and evaluated process placements. The \system{} workflow was quantitatively evaluated on JUWELS Booster.} \label{fig:platform-and-placements} 
\end{figure}

We assessed \system’s measurement and calibration workflow on GPU-accelerated JUWELS Booster system at the Jülich Supercomputing Centre~\cite{jscJuwelsBoosterOverview}, as shown in Figure~\ref{fig:platform-and-placements}. We refer to the two JUWELS Booster topological placement allocations as Placement~A and Placement~B. At 64 GPUs, as depicted in Figure~\ref{fig:placements}, Placement~A crosses two adjacent Dragonfly+ groups, whereas Placement~B remains within one group. The scale-up parameter and scale-out is calibrated using Placement~A. Placement~B is used only for scale-out validation.

\subsection{Benchmarks and Ground-truth Measurements}\label{subsec:ground-truth}
We use NCCL Ring/Simple collective communication benchmark executions with captured channel assignments~\cite{nvidia26nvidia}. The evaluated collective operations are AllGather, AllReduce, and ReduceScatter. All four GPUs of one node participate in the scale-up experiments. The scale-out experiments use 8, 16, 32, or 64 GPUs.

The four-GPU executions use 24 NCCL channels. Executions at larger scales use eight channels. For each collective and GPU count, the dataset contains 1,024 operation sizes between 0.5~MiB and 512~MiB.

Our primary evaluation metric is collective completion time, which is defined as the average in-place operation time reported by \texttt{nccl-tests} over 20 timed iterations following one warm-up iteration. We also track algorithmic bandwidth as defined in Equation~\ref{eq:algorithmic-bandwidth}. Algorithmic bandwidth is derived from completion time and is used to explain the shape of the performance curves.

\subsection{Simulation Execution Engine Instantiation}

\subsubsection{Simulation Framework}
\label{sec:simulator-configuration}
We modeled the Dragonfly+ topology of the JUWELS Booster network in SST/Merlin~\cite{shpiner17dragonflyplus,jscJuwelsBoosterOverview}. The model preserves cell hierarchy, local and global links, link capacities, and measured node placement.

The default ASTRA-sim collective implementation derives a communication pattern from the collective type and communicator size. This is not sufficient when a measured NCCL execution uses a specific communication plan. We therefore modified astrasim to enable it to read the captured NCCL log (i.e., selected algorithm, channels, chunking, GPU-to-NIC placement) and to approximately reproduce the corresponding communication plan.

\begin{figure}[t]
    \centering
    \resizebox{0.7\linewidth}{!}{%
        \begin{tikzpicture}
\begin{axis}[
    width=\linewidth,
    height=0.48\linewidth,
    xlabel={Iteration},
    ylabel={Loss},
    xmin=0, xmax=30,
    ymin=0, ymax=0.37,
    xtick={0,5,10,15,20,25,30},
    grid=major,
]

\addplot[
    mark=none,
    line width=1.5pt,
]
coordinates {
    (0,  0.3541691013)
    (1,  0.2194541286)
    (2,  0.1745491377)
    (3,  0.1520966422)
    (4,  0.1282653644)
    (5,  0.1123778455)
    (6,  0.1010296184)
    (7,  0.0528724434)
    (8,  0.0432052607)
    (9,  0.0335380780)
    (10, 0.0238708953)
    (11, 0.0216035558)
    (12, 0.0190569067)
    (13, 0.0165102577)
    (14, 0.0162309481)
    (15, 0.0159516385)
    (16, 0.0156723289)
    (17, 0.0153930193)
    (18, 0.0150938410)
    (19, 0.0150739722)
    (20, 0.0150541035)
    (21, 0.0150342347)
    (22, 0.0150143659)
    (23, 0.0149944972)
    (24, 0.0149746284)
    (25, 0.0149746284)
    (26, 0.0149746284)
    (27, 0.0149746284)
    (28, 0.0149746284)
    (29, 0.0149746284)
    (30, 0.0149746284)
};

\end{axis}
\end{tikzpicture}
    }
    \caption{completion-time RMSE as function each iteration.}
    \label{fig:loss-convergence}
\end{figure}
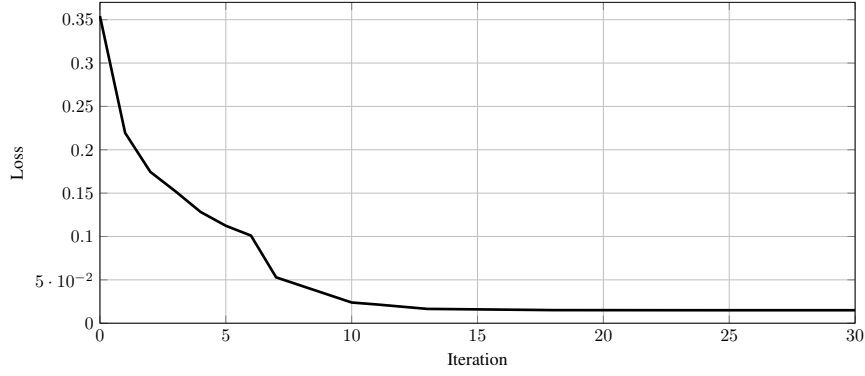

The simulations run on a  server equipped with two Intel Xeon E5-2699 v3 processors (36 physical cores, 72 hardware threads) running Ubuntu 22.04. The optimizer processes are batched sequentially. We parallelize the execution of simulator runs across 72 processes, which results in each iteration taking a median 84.5s of wall-clock time. Each calibration is assigned 30 iterations as a fixed budget (see Figure.~\ref{fig:loss-convergence}), where each iteration can consist of up to 512 independent simulator executions.

\subsubsection{Tracked Simulator Output Metrics}
The simulator output primarily tracked is the completion time of the simulated collective. This is used to determine the loss compared to the ground truth from real hardware measurement, as described in subsection~\ref{subsec:ground-truth}.
For each simulation, the simulator reproduces the recorded GPU placement, GPU-to-NIC
affinity, collective algorithm, protocol, channel count, channel order, and message partitioning. This prevents calibration from compensating for differences between the measured NCCL execution
and the simulated collective communication benchmarks.

\subsubsection{Benchmark Partitioning for Calibration and Validation}
Each collective communication benchmark contains 1,024 operation sizes. During calibration, 389 distinct sizes are utilized while the remaining 635 sizes are used for validation, as shown in Table~\ref{tab:data-split}. For scale-up (i.e. $n=4$) stage of calibration, we use only a subset of the four-GPU AllReduce benchmarks targetting Placement~A topology, whereas the rest of the collectives~(AllGather and ReduceScatter) are entirely used for validation. 


For scale-out calibration (i.e. $n>=8$), we retain the fixed scale-up configuration and use only a subset of the 64-GPU AllReduce measurements from Placement~A. The resulting configuration is validated in three ways. We first evaluate the remaining 64-GPU operation sizes from Placement~A. We then evaluate the complete 8, 16, and 32-GPU AllReduce benchmarks targeting Placement~A to test generalization across communicator size. Finally, we validate the complete 8, 16, 32, and 64-GPU AllReduce benchmarks targeting Placement~B to test generalization across physical allocation, since Placement~B topology was not used during calibration.

\begin{table*}[t]
\caption{Calibration and validation benchmarks used in the scale-up and scale-out
experiments. Calibration uses only AllReduce measurements targeting Placement~A topology.
All other listed measurements are evaluated after the corresponding
configuration has been fixed. $-$ indicates the configuration was not used for calibration, and only employed for validation.}
\label{tab:data-split}
\centering
\scriptsize
\begin{tabular}{
    >{\raggedright\arraybackslash}p{0.11\textwidth}
    >{\raggedright\arraybackslash}p{0.20\textwidth}
    >{\centering\arraybackslash}p{0.10\textwidth}
    >{\centering\arraybackslash}p{0.08\textwidth}
    >{\centering\arraybackslash}p{0.11\textwidth}
    >{\centering\arraybackslash}p{0.12\textwidth}
    >{\centering\arraybackslash}p{0.12\textwidth}
}
\toprule
\textbf{Stage}
& \textbf{Collective}
& \textbf{Placement}
& \textbf{GPUs}
& \textbf{Calibration}
& \textbf{Validation} 
& \textbf{max(RMSE)} \\ 
\midrule

Scale-up
& AllReduce
& A
& 4
& 389
& 635 & 0.0139 \\

Scale-up
& AllGather
& A
& 4
& $-$
& 1,024 & 0.0131\\

Scale-up
& ReduceScatter
& A
& 4
& $-$
& 1,024 & 0.0105 \\

\midrule

Scale-out
& AllReduce,AllGather,ReduceScatter
& A
& 64
& 389
& 635 & 0.0784\\

Scale-out
& AllReduce,AllGather,ReduceScatter
& A
& 8, 16, 32
& $-$
& 1,024 & 0.3349\\

Scale-out
& AllReduce,AllGather,ReduceScatter
& B
& 8, 16, 32, 64
& $-$
& 1,024 & 0.2134 \\

\bottomrule
\end{tabular}
\end{table*}

\begin{table*}[t]
    \centering
    \caption{Representative subset of the most relevant direct (DP)
    and auxiliary (AP) parameters. Checkmark~($\checkmark$) indicates applicability of the corresponding parameter
    to scale-up and scale-out communication. The final value of the DP refers to value set based on target hardware platform whereas for AP, it refers to the value obtained at the end of calibration.}
    \label{tab:final-parameters}
    \small
    \setlength{\tabcolsep}{5pt}
    \resizebox{\textwidth}{!}{%
    \begin{tabular}{@{}cllcccl@{}}
        \toprule
        \textbf{Type} &
        \textbf{Component} &
        \textbf{Parameter} &
        \textbf{Scale-up} &
        \textbf{Scale-out} &
        \textbf{Domain} &
        \textbf{Final value} \\
        \midrule

        \multirow{5}{*}{%
            \rotatebox[origin=c]{90}{%
                \shortstack{%
                    \scriptsize\bfseries Direct\\
                    \scriptsize\bfseries (DP)%
                }%
            }%
        } &
        Scale-up network &
        GPU-to-GPU link bandwidth &
        $\checkmark$ & &
        -- &
        $100$ GB/s per peer \\

        & Scale-out network &
        GPU-to-NIC bandwidth &
        & $\checkmark$ &
        -- &
        $200$ Gbit/s per NIC \\

        & Scale-out network &
        Network-link bandwidth &
        & $\checkmark$ &
        -- &
        $200$ Gbit/s \\

        & Dragonfly+ &
        Intra-group link delay &
        & $\checkmark$ &
        -- &
        $0.5\,\mu$s \\

        & Dragonfly+ &
        Inter-group global-link delay &
        & $\checkmark$ &
        -- &
        $1\,\mu$s \\

        \midrule

        \multirow{5}{*}{%
            \rotatebox[origin=c]{90}{%
                \shortstack{%
                    \scriptsize\bfseries Auxiliary\\
                    \scriptsize\bfseries (AP)%
                }%
            }%
        } &
        Collective &
        Collective startup overhead &
        $\checkmark$ & $\checkmark$ &
        $0$--$300\,\mu$s &
        $32.78\,\mu$s \\

        & Collective &
        Per-loop overhead &
        $\checkmark$ & $\checkmark$ &
        $0$--$50\,\mu$s &
        $6.36\,\mu$s \\

        & Scale-up &
        Per-send scale-up delay &
        $\checkmark$ & &
        $1$--$50\,\mu$s &
        $5.36\,\mu$s \\

        & Scale-out &
        Per-send scale-out delay &
        & $\checkmark$ &
        $0.5$--$30\,\mu$s &
        $18.31\,\mu$s \\

        & Reduction &
        Kernel throughput &
        $\checkmark$ & $\checkmark$ &
        $60$--$300$ GB/s &
        $240.73$ GB/s \\

        \bottomrule
    \end{tabular}%
    }
\end{table*}

\subsection{Calibration Optimizer}
\label{subsec:bayesian_opt}
We choose batch Bayesian optimization implemented with BoTorch~\cite{balandat2020botorch}. A Gaussian-process surrogate models
the relationship between auxiliary parameters and completion-time RMSE. The input parameters are normalized, and the observed losses are standardized.

The default~\footnote{The default configuration uses manually selected auxiliary parameter (AP) values based on our best understanding of the system.} baseline configuration includes points distributed across the allowed parameter space. Subsequent iterations use logarithmic expected improvement~\cite{ament2023unexpected} to propose batches of two configurations. 

\subsection{Scale-up Domain Calibration}

The scale-up configuration is calibrated using only four-GPU AllReduce measurements from Placement~A. Because all four GPUs are located in one node, these measurements expose the scale-up network fabric without introducing scale-out communication. After calibration, all scale-up parameters are fixed.

We first evaluate the calibrated parameter configuration on the 635 AllReduce operation sizes from Placement~A that did not influence calibration. We then apply the same configuration to the complete AllGather and ReduceScatter curves. These collectives were not used during fitting or configuration selection. Finally, we repeat the three collective experiments on Placement~B.

\begin{figure}[!htb]
    \centering
    \includegraphics[width=\textwidth]
    {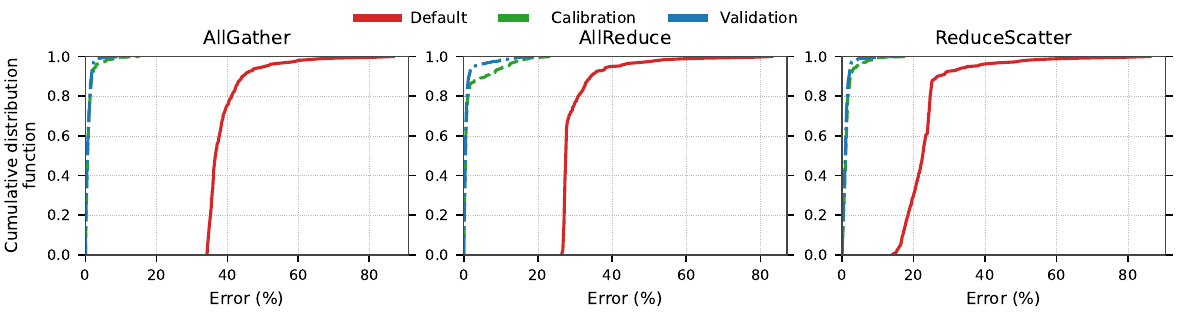}
    \caption{Distribution of absolute relative completion-time error for four GPUs. The simulator is calibrated using only a subset of the AllReduce measurements from Placement~A. AllGather, ReduceScatter, Placement~B, and the remaining AllReduce operation sizes are used only for validation.}
    \label{fig:su-msg-cdf}
\end{figure}

Figure~\ref{fig:su-msg-cdf} compares the error distributions of the default and calibrated configurations. The default configuration produces mean errors of 39.0\% for AllGather, 29.9\% for AllReduce, and 23.5\% for ReduceScatter. Using the configuration calibrated only on AllReduce, the corresponding mean validation errors are 0.77\%, 0.94\%, and 1.05\%.
The 95th-percentile errors are 2.02\%, 2.74\%, and 2.27\%, respectively.

\begin{figure}[!htb]
    \centering
    \includegraphics[width=\textwidth]
    {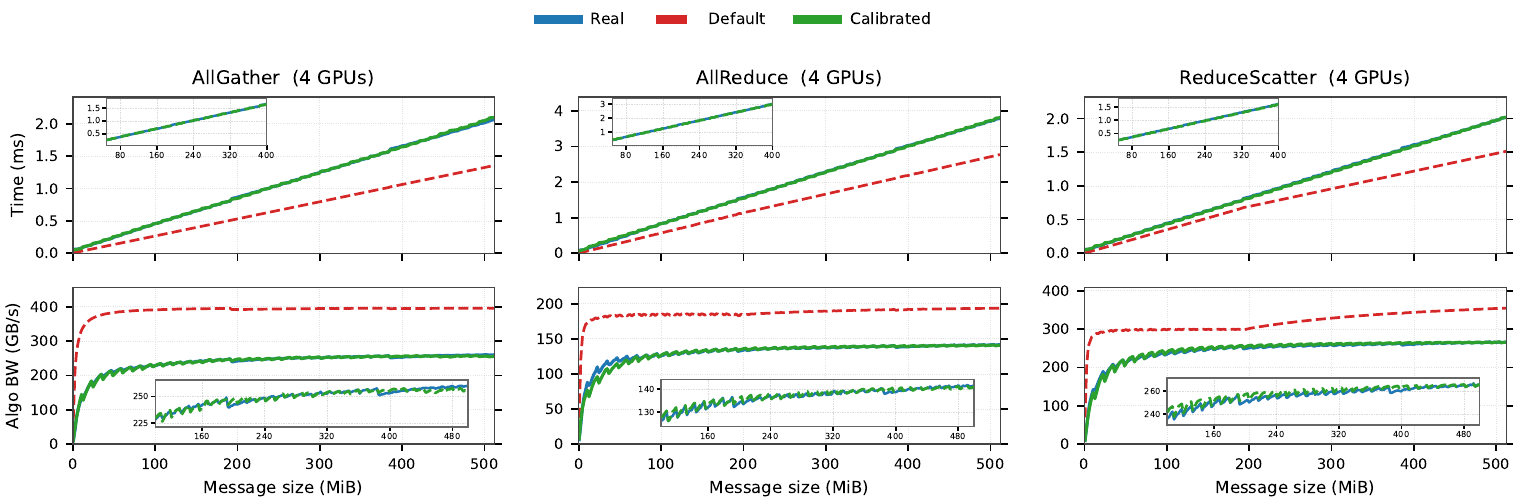}
    \caption{Measured, default, and calibrated scale-up performance for 4
    GPUs. Only AllReduce from Placement~A is used for calibration. The
    AllGather and ReduceScatter results test cross-collective generalization
    without retuning.}
    \label{fig:calib-curves}
\end{figure}

Figure~\ref{fig:calib-curves} shows the measured, default, and calibrated completion-time and algorithmic-bandwidth curves. The calibrated configuration corrects the message size trend and the large message size bandwidth plateau. The same parameter configuration also follows the AllGather and ReduceScatter curves even though neither collective was used during calibration.

\keytakeaway{
Using \system, a subset of collective communication benchmark data (in this case AllReduce) is sufficient to effectively calibrate the evaluated scale-up model. The calibrated parameter configuration is observed to preserve simulation accuracy on other AllReduce operation sizes, two other collectives, and a different placement within the scale-up domain.
}

\subsection{Scale-out Domain Calibration}

After scale-up calibration, the complete set scale-up paramters is fixed. Scale-out calibration uses only a subset of the 64-GPU AllReduce measurements from Placement~A. At this scale, the collective crosses two Dragonfly+ groups.

The collective communication benchmark used for scale-out calibration must expose the parameter being estimated strongly enough to separate it from other costs. A simplified timing expression for ring AllReduce is:

\begin{equation}
\label{eq:ring-identifiability}
    T_{\mathrm{AR}}(n,m)
    \approx
    \alpha_{\mathrm{AR}}
    +2(n-1)\delta_{\mathrm{send}}
    +R(n,m)\delta_{\mathrm{loop}}
    +\frac{V(n,m)}{B}.
\end{equation}

The collective startup term $\alpha_{\mathrm{AR}}$ is paid once, whereas the scale-out send delay $\delta_{\mathrm{send}}$ is paid during each of the $2(n-1)$ ring steps. At small communicator sizes, different combinations of startup and send delay can therefore produce similar completion times. As the communicator grows, the repeated send-delay contribution becomes larger and easier to distinguish from the one-time startup cost. In particular, its contribution is 21 times larger at
64 GPUs than at 4 GPUs. The same reasoning applies to other ring collectives, although AllGather and ReduceScatter each contain $n-1$ communication steps rather than $2(n-1)$.

We therefore estimate the scale-out send delay using the 64-GPU AllReduce execution from Placement~A. As shown in Figure~\ref{fig:x-gpu}, the calibrated 64-GPU configuration achieves a mean absolute relative error of 1.29\% over the 635 held-out message sizes.

\keytakeaway{
Collective communication with large communicator sizes makes repeated per-send costs easier to distinguish from one-time collective startup costs. This reduces the ability of one parameter to compensate for another during calibration.
}

\subsection{Validating different Communicator Sizes and Collectives}

After the scale-out system calibration employing 64 GPUs, the complete parameter configuration for the simulator is finalized. We first validate the efficacy of calibration on system scales of 8, 16, and 32 GPUs from Placement~A. No parameter is retuned for these communicator sizes.

\begin{figure}[!h]
    \centering
    \includegraphics[width=\textwidth]
    {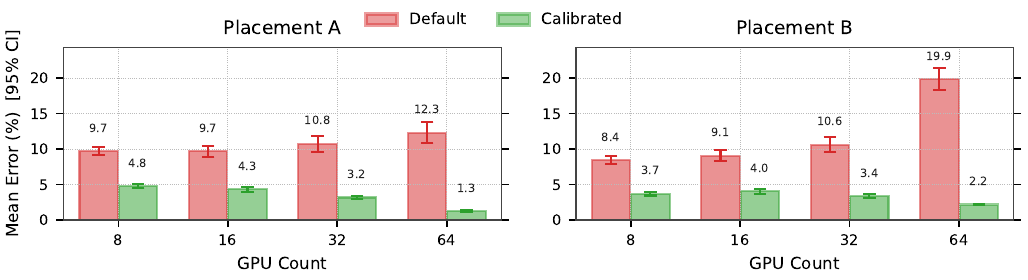}
    \caption{Distribution of AllReduce prediction error across communicator
    sizes on Placement~A. Scale-out calibration uses a subset of the 64-GPU
    measurements. The 8, 16, and 32 GPU system configurations are validated  without parameter retuning.}
    \label{fig:x-gpu}
\end{figure}

Figure~\ref{fig:x-gpu} summarizes the corresponding error distributions for AllReduce. The mean relative errors are 4.77\% at 8 GPUs, 4.34\% at 16 GPUs, and 3.17\% at 32 GPUs. At 64 GPUs, the mean error on operation sizes excluded from calibration is 1.29\%.The relative error for the calibrated parameter configuration is larger at smaller communicator sizes because the calibrated scale-out delay is incurred fewer times. Costs paid once per collective and remaining scale-up errors therefore represent a larger fraction of the total completion time.

\begin{figure}[!h]
    \centering
    \includegraphics[width=\textwidth]
    {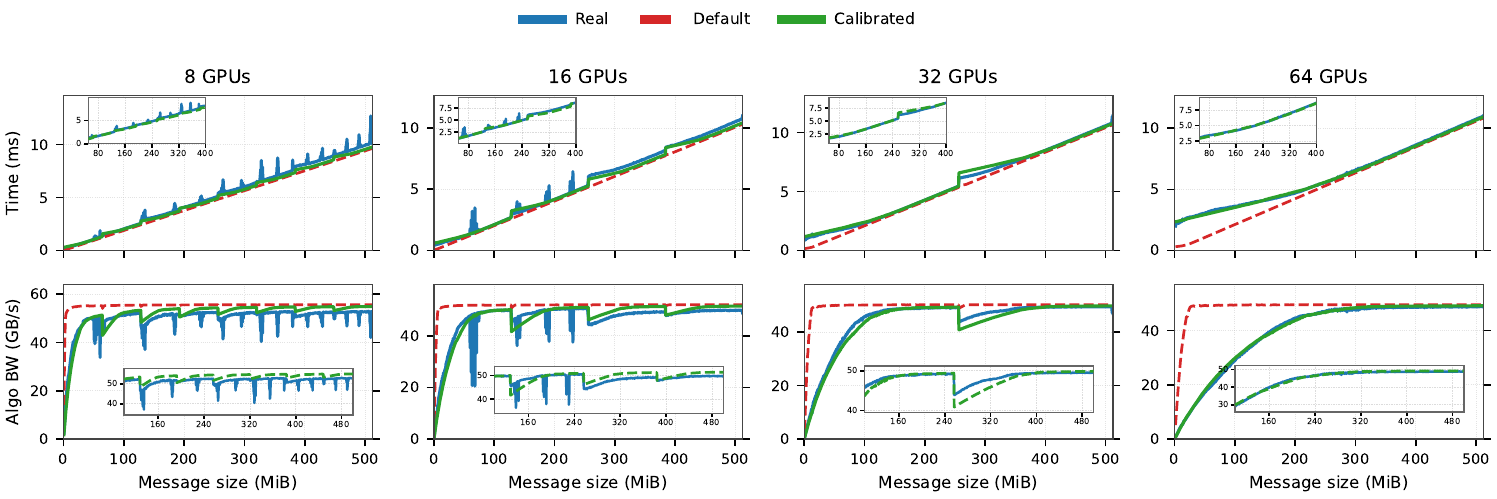}
    \caption{Measured, default, and calibrated AllReduce performance from
    8 to 64 GPUs on Placement~A. Only a subset of the 64-GPU AllReduce curve
    is used for scale-out calibration.}
    \label{fig:xgpu-calib-A}
\end{figure}

Figure~\ref{fig:xgpu-calib-A} compares the measured, default, and calibrated AllReduce results. The calibrated parameter configuration is observed to preserve the trends for completion-time trend and algorithmic-bandwidth across evaluated GPU counts.

\keytakeaway{
Using CALIPER, parameters calibrated only with AllReduce measurement data from one topological placement are observed to retain validation accuracy for different topological placements \& system scales.
}

\section{Discussion}
\label{sec:discussion}


In this section, we discuss the directions that can be taken to extend \system{} methodology and to adopt it to other scenarios than those presented in the previous section.
\subsection{Setting up Calibration Budgets and Stopping Criteria}
In the reported set of experiments, we setup a fixed iteration budget specified at the start of each calibration run. This approach could be extended to a combination of a maximum resource budget~(expressed in number of iterations, simulator execution runs, or wall-clock time), with an early stopping criterion~(in terms of target RMSE) tracking calibration progress. Also,  if the RMSE does not change across a set of multiple consecutive iterations (without reaching the target RMSE), it may be used as a stopping criteria, indicating errors in the simulator which cannot be fixed by calibrating the simulator parameters. 




\subsection{Adapting to different network simulator}


\system{} separates its calibration logic from simulator-specific input files, commands, and output formats through the simulator interface, as shown in the Fig.~\ref{fig:model}. This freedom allows a new network simulator to be calibrated using \system{} by means of an adapter specific to that simulator to attach the simulation execution engine and report the simulator outputs in the requried format.


\subsection{From Collectives to End-to-End Workloads}
Distributed ML training and inference serving simulators rely on network simulators for forecasting performance of AI workloads~\cite{lu17flexflow,
won23astrasim,
ardalani24deepflow,
cho24llmservingsim,
wang25simai,
shen25atlahs,
svedas2026evaluating}. These simulators deal with a large variety of parameters to model including distribution strategy of data, computation time, dependencies between phases, concurrent communication, and overlap between computation and communication. By using the \system{} methodology, these simulators can pre-calibrate their network simulators to represent target hardware systems faithfully, thereby enabling more reliable performance prediction of ML workloads.

\section{Related Work}
\label{sec:related_work}

High-fidelity simulation studies commonly assess accuracy by comparing simulated and observed performance under a selected set of conditions. However, a close fit alone is insufficient evidence that a simulator represents the target system correctly: errors in workload mapping, uncertain hardware parameters, and omissions in the model can compensate for each other and yield accurate predictions for incorrect underlying reasons. 
Prior work has developed principles for simulation credibility and validation~\cite{mehta16SimulationCredibility,pawlikowski02credibility,sargent20VnV,velho13tcp}, techniques for simulator calibration~\cite{asri16simulator,davoudzade24EnhancingNS,mcdonald24calibration,suetterlein24automatic,suetterlein24hpc}, and increasingly sophisticated network and distributed-AI simulators~\cite{gui25parallel,shen25atlahs,wang25simai,won23astrasim}. However, these strands do not consistently distinguish three conceptually different activities: aligning simulator inputs with directly measurable system properties and inferring auxiliary parameters that account for unresolved or abstracted behavior. \system{} makes this separation explicit and uses it to structure the calibration and validation of AI/HPC network simulators.

\noindent\textbf{Simulation credibility, verification, and validation.}
Simulation credibility requires more than a close match between simulated and observed results for a limited set of experiments. Verification and validation distinguish correct implementation from whether a simulator, its assumptions, and its inputs adequately represent the target system~\cite{sargent20VnV,mehta16SimulationCredibility}. In network simulation, concerns about credibility also include the quality of experimental methodology and statistical analysis~\cite{pawlikowski02credibility}, while attempts to invalidate a model can expose flaws that remain hidden when evaluation reports only successful cases~\cite{velho13tcp}. \system{} translates these general principles into a concrete workflow for AI/HPC network simulators: it separates hardware alignment, parameter inference, and model diagnosis, then uses structured validation on withheld workloads and scales to test whether a calibrated configuration is credible beyond the observations used for fitting.

\noindent\textbf{Automated calibration.}
Automated calibration reduces the effort of manually selecting simulator parameters by searching for configurations that minimize the difference between simulated and measured behavior~\cite{mcdonald24calibration}. This is valuable because calibration becomes increasingly difficult as the number of parameters and their interactions grow~\cite{hofmann05complexity}. Techniques such as Bayesian optimization have been used to tune HPC network models across message regimes~\cite{suetterlein24automatic,suetterlein24hpc}, while other work combines calibration with network model optimization and evaluates configurations across message lengths and router models~\cite{davoudzade24EnhancingNS}. However, automatic calibration has also shown that an inaccurate baseline can substantially distort prediction quality~\cite{asri16simulator}. \system{} differs by treating optimization as one stage of a broader method. It separates parameters that should be measured directly from those that must be inferred, constrains inferred values using system knowledge, and uses independent validation to determine whether residual error comes from parameter choices, workload mapping, or missing model behavior.

\noindent\textbf{Simulation accuracy and transferability.}
Simulation accuracy depends on both the chosen network abstraction and the communication conditions being modeled. Flow level models can support faster simulation than packet level models, but their accuracy varies across operating regimes and may become unacceptable when their abstraction omits relevant behavior~\cite{fujiwara07speed}. Incorporating network topology and contention can improve predictive accuracy across real application benchmarks~\cite{bedaride14mpi}, while broad evaluations of distributed system simulators show that accuracy and scalability must be assessed across various different workloads and platforms rather than assumed from a single setting~\cite{casanova14simgrid, monniot24simulation}. \system{} addresses transferability by comparing cross collective, and scale generalizing calibration strategies, then evaluating each resulting configuration on workloads and scales that were not used during fitting.

\section{Conclusion}
\label{sec:conclusion}

We presented \system, a principled methodology for calibrating AI/HPC network simulators that moves beyond fitting simulation outputs to hardware measurements. \system{} explicitly separates direct parameter alignment, auxiliary-parameter inference, and independent validation, preventing optimization from compensating for known system properties, workload mismatches, or model-form limitations. Our evaluation demonstrates the importance of identifiability-aware calibration, where network domains are calibrated hierarchically and calibration experiments are selected to sufficiently expose the parameters being inferred. By calibrating the simulator configuration and evaluating it across held-out message sizes, system scales, and physical placements, \system{} distinguishes goodness-of-fit from predictive validity. More broadly, our results show that simulator credibility cannot be established by low calibration error alone: it requires evidence that the inferred configuration captures system behavior that persists beyond the observations used for fitting. \system{} provides a systematic foundation for obtaining such evidence and for making simulator calibration more interpretable and scientifically defensible.


\bibliographystyle{unsrtnat}
\bibliography{refs}

\end{document}